\documentclass[12pt]{article}
\usepackage{fullpage,cmu-titlepage2}
\usepackage{times}

\usepackage{amsmath}
\usepackage{rotating}
\usepackage{tabularx}
\usepackage[
pageanchor=true,
plainpages=false,
pdfpagelabels, 
bookmarks,
bookmarksnumbered,
pdfborder=0 0 0,
pdfpagemode=UseOutlines]{hyperref}
\usepackage{draftstamp}
\usepackage{subfigure}

\usepackage{times}
\usepackage{amssymb}
\usepackage{graphicx}
\usepackage{wrapfig}
\usepackage{algorithmic,algorithm}
\usepackage{url}

\newcommand{\Pb}{\mathbb{P}}
\newcommand{\zv}{\mathbf{z}}
\newcommand{\sv}{\mathbf{s}}
\newcommand{\wv}{\mathbf{w}}
\newcommand{\ev}{\mathbf{e}}
\newcommand{\fv}{\mathbf{f}}
\newcommand{\yv}{\mathbf{y}}

\newcommand{\SMF}[0]{SM\ensuremath{^{\textrm{4}}}}

\title{Understanding the Interaction between Interests, Conversations and Friendships in Facebook}

\author{Qirong Ho \and Rong Yan$^{\ast}$ \and Rajat Raina$^{\ast}$ \and Eric P. Xing}

\date{November 2012}

\abstract{
In this paper, we explore salient questions about user interests, conversations and friendships in the Facebook social network,
using a novel latent space model that integrates several data types. A key challenge of studying Facebook's data is the
wide range of data modalities such as text, network links, and categorical labels.
Our latent space model seamlessly combines all three data modalities over millions of users, allowing us to study the
interplay between user friendships, interests, and higher-order network-wide social trends on Facebook.
The recovered insights not only answer our initial questions, but also reveal surprising facts about user interests
in the context of Facebook's ecosystem. We also confirm that our results are significant with
respect to evidential information from the study subjects.
}


\trnumber{CMU-ML-12-109}

\citationinfo{
$^{\ast}$Facebook, 10 Hacker Way, Menlo Park CA 94025 \\
}

\begin{document}
\renewcommand*{\thepage}{title-\arabic{page}} 
\maketitle
\renewcommand*{\thepage}{\arabic{page}}

\section{Introduction}
From blogs to social networks to video-sharing sites and still others,
online social media have grown dramatically over the past half-decade. These media host and aggregate
information for hundreds of millions of users, and this has sired an unprecedented opportunity to study people on an incredible scale,
and over a broad spectrum of open problems. In particular, the study of user interests, conversations and friendships is of special value
to the health of a social network ecosystem. As a classic example,
if we had a good guess as to what a user likes (say, from explicit labels or conversations), we could serve her more appropriate content,
which may increase her engagement with the media, and potentially help to obtain more structured data about her interests.
Moreover, by providing content that is relevant to the user {\it and her friends}, the social network can increase engagement
beyond mere individual content consumption --- witness the explosive
success of social games, in which players are rewarded for engaging in game activities with friends, as opposed to solitary play.

These examples illustrate how social networks depend on the interplay between user interests, conversations and friendships.
In light of this, we seek to answer several questions about Facebook:
\small
\begin{list}{\labelitemi}{\leftmargin=1em}
\item
How does Facebook's social (friendship) graph interact with its interest graph and conversational content? Are they correlated?
\item
What friendship patterns occur between users with similar interests?
\item
Do users with similar interests talk about the same things?
\item
How do different interests (say, camping and movies) compare? Do groups of users with distinct interests also exhibit
different friendship and conversational patterns?
\end{list}
\normalsize
To answer these questions on the scales dictated by Facebook, it is vital to develop tools that can
visualize and summarize user information in a salient and aggregated way over large and diverse populations of users.
In particular, it is critical that these tools enable macroscopic-level study of social network phenomena, for there are simply too many
individuals to study at fine detail. Through the lens of these tools, we can gain an understanding of how user interests, conversations
and friendships make a social network {\it unique}, and how they make it {\it function}. In turn, this can shape policies aimed at retaining
the special character of the network, or at enabling novel utilities to drive growth.

\subsection{Key Challenges}

Much research has been invested in user interest {\it prediction}
\cite{blei2008supervised,bell2007lessons,mooney2000content,hill1995recommending,basu2001technical},
particularly methods that predict user interests by looking at similar users.
However, existing works are mostly built on an incomplete view of the social media data, often solely restricted to user texts.
In particular, the network itself acts a conduit for information flow among users, and we cannot attain a complete view of the social media
by ignoring it. Thus, a deep, holistic understanding of 
user interests and of the network as a whole requires a perspective over diverse data modalities (views) such as text, network links and categorical labels. To the best of our knowledge, a principled approach that enables such capability has yet to be developed. Hence, our goal is to produce such a system for understanding the relationships between user interests, conversations and friendships.

In developing this system, at least two challenges must be properly addressed.
For one, the data scale is unprecedented ---
Facebook has hundreds of millions of active users, with diverse modalities of
information associated their profiles: textual status updates, comments on other user's pages, pictures, and friendships,
to name a few. Any method that does not
scale linearly in the amount of data is bound to fail.
The other challenge is the presence of complex {\it structure} in
Facebook's data; its information is not presented as a simple feature vector,
but as a cornucopia of structured inputs, multimodal in the sense that text, networks, and label
data each seemingly requires a different approach to learning. Even the text alone cannot be treated as a simple
bag of words, for it is separated into many comments and posts, with potentially sharp changes of topics and intents.
One cannot fully model this rich structure with methods that require user data to be input as flat feature vectors,
or that require a similarity function between them.


\subsection{Solutions}

With these challenges in mind, we present a scalable machine learning system that we use to visualize and explore
the interests of millions of users on Facebook, and that potentially scales to tens or hundreds of millions of users.
The key to this system is a unified latent space model jointly over text, network and label data,
where some of its building blocks have been inspired by earlier successful attempts on certain modalities, such as the supervised Latent Dirichlet
Allocation model over text and labels \cite{blei2008supervised}, the Mixed Membership Stochastic Blockmodel
over networks \cite{airoldi2008mixed}, and the joint text/citation topic models of Nallapati {\it et al.}~\cite{nallapati2008joint}.
We call our model the Supervised Multi-view Mixed Membership Model (\SMF),
which surmounts the multimodal data challenge by transforming user text, network and label data into an integrated latent feature vector for each user,
and overcomes the scalability challenge by first training model parameters on a smaller subset of data, after which it infers
millions of user feature vectors in parallel. Both the initial training phase and the integrated feature vector inference
phase require only linear time and a single pass through the data.

Our system's most important function is visualization and exploration, which is achieved by deriving other kinds of information from the data in a principled, statistical manner.
For instance, we can summarize the textual data as collections of related words,
known as {\it topics} in the topic modeling literature~\cite{blei2008supervised,blei2009topic}. Usually, these topics will be coherent
enough that we can assign them an intuitive description, e.g. a topic with the words ``basketball", ``football" and ``baseball"
is best described as a ``sports" topic.
Next, similar to Blei {\it et al.}~\cite{blei2008supervised}, we can also report the correlation between each topic and the label
under study --- for instance, if we are studying the label ``I vote Democratic", we would expect topics containing the
words ``liberal" and ``welfare" to be positively correlated with said label. The value of this lies in finding {\it unexpected}
topics that are correlated with the label. In fact, we will show that on Facebook, certain well-known brands are positively correlated with generic
interests such as movies and cooking, while social gaming by contrast is negatively correlated.
Finally, we can explain each friendship
in the social network in terms of two topics, one associated with each friend. The motivation behind this last feature
is simple: if we have two friends who mostly talk about sports, we would naturally guess that their friendship is due to mutual interest in sports.
In particular, interests with a high degree of mutual interest friendships are valuable from a friendship recommendation perspective. As an example,
perhaps ``sports" is highly associated with mutual interest friendships, but not ``driving". When ranking potential friends for a user who likes sports and driving,
we should prefer friends that like sports over friends that like driving, as friendships could be more likely to form over sports.

From this latent topical model, we can construct visualizations like Figure \ref{fig:interest_viz} that summarize
all text, network and label data in a single diagram.
Using this visualization, we proceed with the main application of this paper, a cross-study of four general user interests, namely ``camping", ``cooking", ``movies", and ``sports".
Our goal is to answer the questions posed earlier about user interests, conversations
and friendships in Facebook, and thus glean insight into what makes Facebook unique, and how it functions.
We also justify our analyses with quantitative results: by training a linear classifier~\cite{fan2008liblinear} on the four interest labels and our system's user feature vectors,
we demonstrate a statistically significant improvement in prediction accuracy over a bag-of-words baseline.

\begin{figure*}
	\begin{center}
	\includegraphics[width=\textwidth]{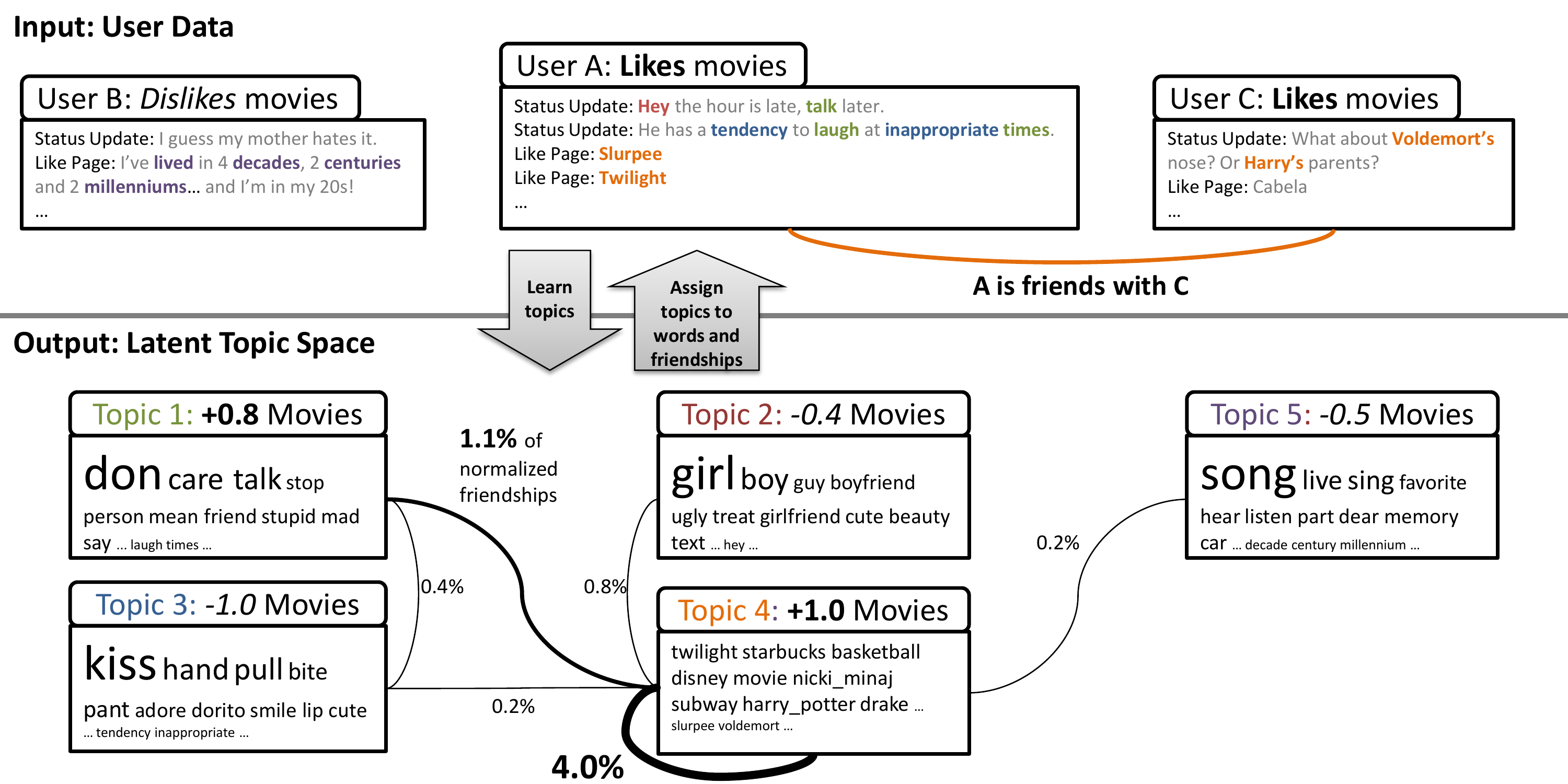}
	\end{center}
	\caption{\small From user data to latent topic space, and back (best viewed in color). User data in the form of text (status updates and like page titles), friendships
	and interest labels (e.g. likes/dislikes movies) is used to learn a latent space of topics. Topics are characterized by a set of weighted keywords,
	a positive or negative correlation with the interest (e.g $+1.0$ Movies), and topic-topic friendship probabilities (expressed as the
	percentage of observed friendships, normalized by topic popularity). After learning the topics, we can assign the most probable topic to
	each user word, as well as the most probable topic-pair to each friendship --- these assignments are represented by word and link colors.
	Observe that users with lots of green/orange words/friendships are likely to be interested in movies, as the corresponding topics (1,4) are detected
	as positive for movies.}
	\label{fig:lsm}
\end{figure*}

\section{Algorithm Overview}
\label{sec:system_overview}
Our goal is to analyze Facebook user data in the context of a general concept, such as ``movies" or ``cooking".
Each Facebook user is associated with three types of data: text such as (but not limited to)
user ``status updates", network links between users based on friendships, and binary labels denoting
interest in the concept (``I like movies") or lack thereof (``I don't like movies"). Intuitively, we want to capture
the relationship between concepts, user text and friendships: for a given concept, we seek words correlated with interest
in that concept (e.g. talking about actors may be correlated with interest in movies), as well as words that are most frequently associated
with each friendship (e.g. we might find two friends that often talk about actors). By learning and visualizing such relationships between the input text, network
and label data (see Figure \ref{fig:lsm}), we can glean insight into the nature of Facebook's social structure.

Combining text and network data poses special challenges: while text is organized into multiple documents per user,
networks are instead {\it relational} and therefore incompatible with feature-based learning
algorithms. We solve this using an algorithm that learns a {\it latent feature space} over
text, network and label data, which we call \SMF. The \SMF~algorithm involves the following stages:
\small
\begin{enumerate}
\itemsep -1pt
	\item
	Train the \SMF~probabilistic model on a subset of user text, network and label data. This learns parameters for a $K$-dimensional
	latent feature space over text, network and labels, where each feature dimension represents a ``topic".
	\item
	With these parameters, we find the best feature space representations of all users' text, network and label data.
	For each user, we infer a $K$-dimensional feature vector, representing her tendency towards each of the $K$ topics.
	\item
	The inferred user features have many uses, such as (1) finding which topics are most associated with friendships,
	and (2) training a classifier for predicting user labels.
\end{enumerate}
\normalsize
The feature space consists of $K$ topics, representing concepts and communities that anchor user conversations, friendships and interests.
Each topic has three components: a vector of word probabilities, a vector of friendship probabilities to each of the $K$ topics, and a scalar correlation
w.r.t the user labels. As an example, we might have a topic with the frequent words ``baseball" and ``basketball", where
this topic has a high self-friendship probability, as well as a high correlation with the positive user label ``I like sports". Based on this topic's most frequent words, we might give
it the name ``American sports"; thus, we say that users who often talk about ``baseball" and ``basketball" are talking about ``American sports".
In addition, the high self-friendship probability of the ``American sports" topic implies that such users are likely to be friends, while the high label correlation
implies that such users like sports in general. Note that topics can have high friendship probabilities to {\it other} topics, e.g. we might find
that ``American sports" has a high friendship probability with a ``Restaurants and bars" topic containing words such as ``beer", ``grill" and ``television".

\section{Supervised Multi-View Mixed Membership Model (\SMF)}
\label{sec:generative_process}

\begin{figure}[t]
	\begin{center}
	\includegraphics[width=.4\textwidth]{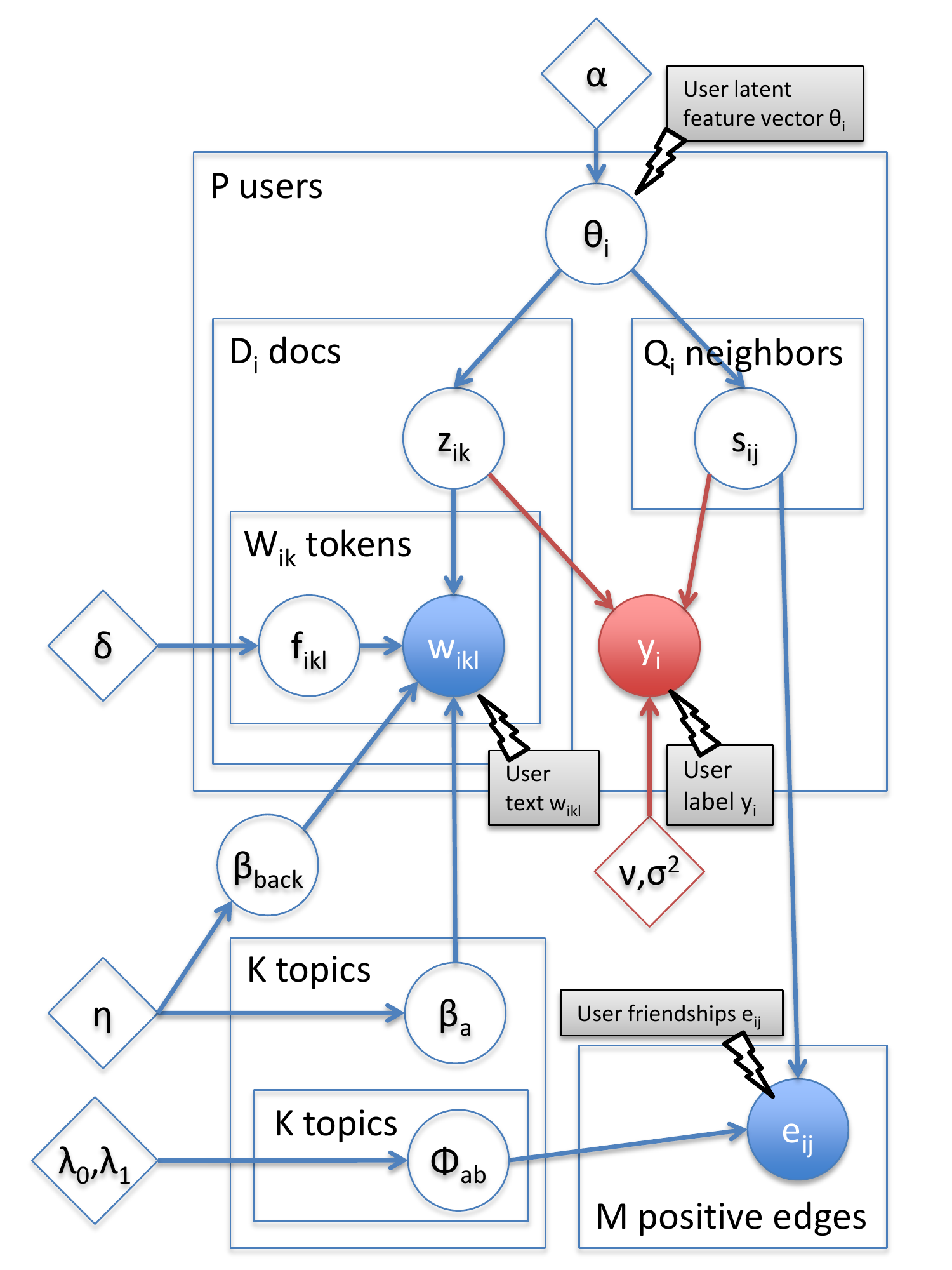}
	\end{center}
	\caption{\small Graphical model representation of \SMF. Tuning parameters are
	diamonds, latent variables are hollow circles, and observed variables
	are filled circles. Variables pertaining to labels $y_i$ are shown in red.}
	\label{fig:graphical_model}
\end{figure}

Formally, \SMF~can be described in terms of a {\it probabilistic generative process}, whose
dependencies are summarized in a graphical model representation (Figure \ref{fig:graphical_model}).
Let $P$ be the number of users, $V$ the text vocabulary size, and $K$ the desired number
of topics. Also let $D_i$ be the number of documents for user $i$, and $W_{ik}$ the number of words in user $i$'s $k$-th document.
The generative details are described below:
\small
\begin{list}{\labelitemi}{\leftmargin=1em}
\setlength{\parskip}{0pt}
	\item
	{\bf Topic parameters:}
	\begin{list}{\labelitemi}{\leftmargin=1em}
		\item
		For the background vocabulary $\beta_{back}$, draw:
		\begin{list}{\labelitemi}{\leftmargin=1em}
			\item
			$V$-dim. word distribution $\beta_{back} \sim \mathrm{Dirichlet}(\eta)$
		\end{list}
		\item
		For each topic $a\in\{1,\dots,K\}$, draw:
		\begin{list}{\labelitemi}{\leftmargin=1em}
			\item
			$V$-dim. topic word distribution $\beta_{a\cdot} \sim \mathrm{Dirichlet}(\eta)$
		\end{list}
		\item
		For each topic pair $(a,b)\in\{1,\dots,K\}^2$, $a\le b$, draw:
		\begin{list}{\labelitemi}{\leftmargin=1em}
			\item
			Topic-topic link probability $\Phi_{ab} \sim \mathrm{Beta}(\lambda_1,\lambda_0)$
		\end{list}
	\end{list}
	\item
	{\bf User features:} For each user $i\in\{1,\dots,P\}$, draw:
	\begin{list}{\labelitemi}{\leftmargin=1em}
		\item
		User feature vector $\theta_i \sim \mathrm{Dirichlet}(\alpha)$
	\end{list}
	\item
	{\bf Text:} For each user document $(i,k) \in \{1,\dots,P\} \times \{1,\dots,D_i\}$:
	\begin{list}{\labelitemi}{\leftmargin=1em}
		\item
		Draw document topic $z_{ik} \sim \mathrm{Discrete}(\theta_i)$
		\item
		For each word $\ell\in\{1,\dots,W_{ik}\}$, draw:
		\begin{list}{\labelitemi}{\leftmargin=1em}
			\item
			Foreground-background indicator $f_{ik\ell} \sim \mathrm{Bernoulli}(\delta)$
			\item
			Word $w_{ik\ell} \sim \mathrm{Discrete}( (\beta_{z_{ik}})^{f_{ik\ell}} (\beta_{back})^{1-f_{ik\ell}} )$
		\end{list}
	\end{list}
	\item
	{\bf Friendship Links:} For each $(i,j)\in\mathrm{EdgeList}$, $i<j$, draw:
	\begin{list}{\labelitemi}{\leftmargin=1em}
		\item
		User $i$'s topic when befriending user $j$, $s_{ij} \sim \mathrm{Discrete}(\theta_i)$
		\item
		User $j$'s topic when befriending user $i$, $s_{ji} \sim \mathrm{Discrete}(\theta_j)$
		\item
		Link $e_{ij} \!\sim\! \mathrm{Bernoulli} (\Phi_{s_{ij},s_{ji}})$ if $s_{ij}\!\le\! s_{ji}$,
		else $e_{ij} \!\sim\! \mathrm{Bern.} (\Phi_{s_{ji},s_{ij}})$
	\end{list}
	\item
	{\bf Labels:} For each user $i\in\{1,\dots,P\}$, draw:
	\begin{list}{\labelitemi}{\leftmargin=1em}
		\item
		Label $y_i \sim \mathrm{Normal}(\hat{\theta}^{\top}_{i} \nu,\sigma^2)$, where
		$\hat{\theta}_{i} = \frac{\sum_k z_{ik} + \sum_j s_{ij}}{D_i + |\mathrm{Neighbors}(i)|}$
	\end{list}
\end{list}
\normalsize
While this generative process may seem complicated at first glance, we shall argue that each component
is necessary for proper modeling of the text, network and label data. Additionally,
the model's complexity does not entail a high runtime --- in fact, our \SMF~algorithm runs in linear time
with respect to the data, as we will show.

\paragraph{Topics and user data}
Each user $i$ has 3 data types: text data $w_i$, network links
$e_{ij}$, and interest labels $y_i\in\{+1,-1\}$. In order to learn salient facts about all 3 datatypes seamlessly,
we introduce a latent space feature vector for each user $i$, denoted by
$\theta_i=(\theta_{i1},\dots,\theta_{iK})$. Briefly, a high value of $\theta_{ia}$ indicates
that user $i$'s text $w_i$, friendship patterns $e_i$ and label $y_i$ are similar to topic $a$.

Every topic $a\in\{1,\dots,K\}$ is associated with 3 objects:
(1) a $V$-dim. word probability vector $\beta_a$,
(2) link formation probabilities $\Phi_{ab}\in[0,1]$ to each of the $K$ topics
$b$, and (3) a coefficient $\nu_a$ that models the linear
dependence of labels $y_i$ with topic $a$.
The vector $\beta_a$ shows
which words are most salient for the topic, e.g. a ``US politics" topic should have high probabilities
on the words ``Republican" and ``Democrat". The link probabilities $\Phi_{ab}$ represent how likely
users talking about topic $a$ are friends with users talking about topic $b$, e.g. ``American sports" having
many friendships with ``Restaurants and bars". Finally, the coefficients $\nu_a$ show the correlation
between topic $a$ and the user interest labels $y_i$.

\paragraph{Text model}
We partition user text data $w_i$ into $D_i$ documents $\{w_{i,1},\dots,w_{i,D_i}\}$, where each doc $ik$
is a vector of $W_{ik}$ words $(w_{ik,1},\dots,w_{ik,W_{ik}})$.
Each document represents a ``status update" by the user, or the title of a page she ``likes".
Compared to other forms of textual data like blogs, Facebook documents are very short.
Hence, we assume each document corresponds to {\it exactly one topic} $z_{ik}$,
and draw all its words $w_{ik\ell}$ from the topic word distribution $\beta_{z_{ik}}$ ---
a notable departure from most topic models \cite{blei2008supervised,chang2009relational}, which
are tailored for longer documents such as academic papers.

Moreover, Facebook documents contain many keywords irrelevant to the main topic.
For example, the message ``I'm watching football with Jim, enjoying it" is about
sports, but the words ``watching" and ``with" are not sports-related.
To prevent such generic words from influencing topic word distributions $\beta_a$,
we introduce per-word foreground-background boolean indicators $f_{ik\ell}\sim\mathrm{Bernoulli}(\delta)$,
such that we draw $w_{ik\ell}$ from $\beta_{z_{ik}}$ as usual when $f_{ik\ell}=1$, otherwise we draw $w_{ik\ell}$
from a ``background" distribution $\beta_{back}$. By relegating irrelevant words to a background
distribution, we can assign topics to entire documents without diluting the topic word distributions with generic words.
More generally, the idea of having separate {\it classes} of word distributions was explored in \cite{zhai2002two,griffiths2005integrating}.

\paragraph{Network model}
Let $\mathrm{Neighbors(i)}$ denote user $i$'s friends, and let $\mathrm{EdgeList}$
denote all friendships $(i,j)$ for $i<j$. Also, let $e_{ij}\in\{0,1\}$ be the adjacency matrix of
friendships, where $e_{ij}=1$ implies $(i,j)\in\mathrm{Edgelist}$. In our model, friendships
arise as follows: first, users $i,j$ draw topics $s_{ij}$ and $s_{ji}$ from their feature vectors $\theta_i,\theta_j$.
Then, the friendship outcome $e_{ij}$ is generated from $s_{ij},s_{ji}$ --- this is in
contrast to words $w_{ik\ell}$, which are generated from only one topic $z_{ik}$. Specifically,
$e_{ij}$ is drawn from a upper-triangular $K\times K$ matrix of Bernoulli parameters $\Phi$; we draw $e_{ij}$
from $\Phi_{s_{ij},s_{ji}}$ if $s_{ij}<s_{ji}$,
otherwise we draw from $\Phi_{s_{ji},s_{ij}}$.
Essentially, $\Phi$ describes friendship probabilities between topics.

Because the Facebook network is {\it sparse}, we only model positive links;
the variables $s_{ij},s_{ji},e_{ij}$ exist if and only if
$e_{ij}=1$. The zero links $e_{ij}=0$ are used in a Bayesian fashion: we
put a $\mathrm{Beta}(\lambda_1,\lambda_0)$ prior on each element of $\Phi$,
and set $\lambda_0 = \ln(\#[\text{zero links}] / K^2)$ and $\lambda_1 = 0.1$,
where $\#[\text{zero links}] =P(P-1)/2 - |\mathrm{EdgeList}|$.
Thus, we account for evidence from zero links without explicitly modeling them,
which saves a tremendous amount of computation.

\paragraph{Label model}
We extract labels $y_i\!\in\!\{+1,-1\}$ from users' ``liked" pages, e.g. ``music" and ``cooking".
By including labels, we can learn which topics are positively/negatively correlated with
user interests. Similar to sLDA \cite{blei2008supervised},
we draw user labels $y_i \!\sim\! \mathrm{Normal}(\hat{\theta}^{\top}_{i}\nu,\sigma^2)$,
where $\hat{\theta}_{i}$ is the average over user $i$'s text topic indicators $z_{ik}$
and network indicators $s_{ij}$ (represented as {\it indicator vectors}). Put simply,
a user's label is a linear regression over her topic vector $\theta_i$.

\subsection{Training Algorithm}

Our \SMF~system proceeds in two phases: a training phase to estimate the latent space topic parameters $\beta,\Phi,\nu,\sigma^2$ from a smaller subset of users, followed by
a parallel prediction phase to estimate user feature vectors $\theta_i$ and friendship topic-pair assignments
$s_{ij},s_{ji}$ for each friendship $e_{ij}=1$. In particular, the $s_{ij},s_{ji}$ provide the most likely ``explanation" for each
friendship, and this forms a cornerstone of our data analysis in Section \ref{sec:user_interest_study}.

Right now, we shall focus on the details of the training algorithm.
Our first step is to simplify the training problem by reducing the number of latent variables, through analytic integration of user feature vectors $\theta$ and
topic word/link parameters $\beta,\Phi$ via Dirichlet-Multinomial and Beta-Binomial conjugacy.
Hence, the only random variables that remain to be inferred are $\zv,\fv,\sv$ (which now depend on
the tuning parameters $\alpha,\eta,\delta$). Once $\zv,\fv,\sv$ have been inferred, we can recover
the topic parameters $\beta,\Phi$ from their values.
We also show that our algorithm runs in linear time w.r.t the amount of data, ensuring scalability.

Training Algorithm (\ref{alg:sm4_training}) alternates between Gibbs sampling on $\zv,\fv,\sv$,
Metropolis-Hastings on tuning parameters $\alpha,\eta,\delta$, and direct maximization of $\nu,\sigma^2$.
This hybrid approach is motivated by simplicity --- Gibbs samplers for models like ours
\cite{griffiths2004finding} are easier to derive and implement than alternatives such as variational inference,
while $\alpha,\eta,\delta$ are easily optimized through the Metropolis-Hastings algorithm.
As for the Gaussian parameters $\nu,\sigma^2$, the high dimensionality of $\nu$ makes MCMC convergence
difficult, so we resort to a direct maximization strategy similar to sLDA \cite{blei2008supervised}.

\subsubsection{Gibbs sampler for latent variables $\zv,\fv,\sv$}

\paragraph{Document topic indicators $\zv$}
A Gibbs sampler samples every latent variable, conditioned on the current values
of all other varibles. We start by deriving the conditional distribution of $z_{ik}$:
\small
\begin{flalign}
	&\Pb(z_{ik} = m \mid \zv_{-ik}, \fv, \wv, \sv, \ev, \yv) \label{eq:gs_z} \\
	&\propto
		\Pb(y_{i} \mid z_{ik} = m, \zv_{i,-k}, \sv_{i})
		\Pb(w_{ik\cdot} \mid z_{ik} = m, \zv_{-ik}, \fv_{ik\cdot}, \wv_{-ik\cdot}) \notag \\
	&\quad \times \Pb(z_{ik} = m\mid \zv_{i,-k}, \sv_{i}) \notag \\
	&\propto
		\exp\left\{-\frac{(y_i - \hat{\theta}^{\top}_{i}\nu)^2}{2\sigma^2}\right\}
		\frac
		{\Gamma(V\eta + \sum_{v=1}^{V} A_v)}
		{\prod_{v=1}^{V} \Gamma(\eta + A_v)}
		\frac
		{\prod_{v=1}^{V} \Gamma(\eta + B_v + A_v)}
		{\Gamma(V\eta + \sum_{v=1}^{V} B_v + A_v)} \notag \\
	&\quad \times
		\left( \#[\{\zv_{i,-k},\sv_{i}\} = m] + \alpha \right), \notag
\end{flalign}
\normalsize
where we use the fact that $\Pb(w_{ik\ell} \mid z_{ik} = m, f_{ik\ell} = 0, \zv_{-ik}, \wv_{-ik\cdot})$ is independent of $z_{ik}$, and where we define
\small
\begin{flalign*}
	&A_v = |\{ (x,y,u) \mid (x,y)\ne(i,k) \,\wedge\, f_{xyu}=1 \,\wedge\, z_{xy}=m \,\wedge\, w_{xyu}=v \}|, \\
	&B_v = |\{ u \mid f_{iku} = 1 \,\wedge\, w_{iku} = v\}|,
\end{flalign*}
\normalsize
where $A_v$ is the number of non-background words $\wv=v$ assigned to topic $m$ and not belonging to user $i$ and document $k$,
and $B_v$ is similar but for words belonging to user/document $ik$. Note that $\hat{\theta}_{i}$ in the
$\exp$ is a function of $z_{ik}$, and was defined in Section \ref{sec:generative_process}.

The distribution of $z_{ik}$ is composed of  a prior term for $z_{ik}=m$ and two posterior terms,
one for user $i$'s label $y_i$, and one for document $ik$'s words $w_{ik\cdot}$. The posterior term for $y_i$ is a Gaussian,
while the posterior term for $w_{ik\cdot}$ is a Dirichlet Compound Multinomial (DCM) distribution, which results
from integrating the word distribution $\beta_{m}$. Notice that background words, i.e.
$w_{ik\ell}$ such that $f_{ik\ell}=0$, do not show up in this posterior term. Finally, the $z_{ik}$ prior term is the DCM
from integrating the feature vector $\theta_i$.

Importantly, the counts $A_v,B_v$ can be cached and updated in constant time for each $z_{ik}$ being sampled, and
therefore Eq. (\ref{eq:gs_z}) can be computed in constant time w.r.t. the number of documents. Hence, sampling all $\zv$ takes linear time in the number of documents.

\paragraph{Word foreground-background indicators $\fv$}
The conditional distribution of $f_{ik\ell}$ is
\small
\begin{flalign}
	&\Pb(f_{ik\ell} = 1 \mid \zv, \fv_{-ik\ell}, \wv, \sv, \ev, \yv) \label{eq:gs_f} \\
	&=
		\Pb(w_{ik\ell} \mid \zv, f_{ik\ell} = 1, \fv_{-ik\ell}, \wv_{-ik\ell}) \Pb(f_{ik\ell} = 1) \notag\\
	&\quad\times
		[\Pb(w_{ik\ell} \mid \zv, f_{ik\ell} = 1, \fv_{-ik\ell}, \wv_{-ik\ell}) \Pb(f_{ik\ell} = 1) \notag\\
	&\qquad+ \Pb(w_{ik\ell} \mid \zv, f_{ik\ell} = 0, \fv_{-ik\ell}, \wv_{-ik\ell}) \Pb(f_{ik\ell} = 0)]^{-1} \notag \\
	&=
		\left(
		\frac{(\eta + E_{w_{ik\ell}})\delta}{V\eta + \sum_{v=1}^{V} E_v}
		\right)
		\left(
		\frac{(\eta + E_{w_{ik\ell}})\delta}{V\eta + \sum_{v=1}^{V} E_v}
		+ \frac{(\eta + F_{w_{ik\ell}})(1-\delta)}{V\eta + \sum_{v=1}^{V} F_v}
		\right)^{-1}, \notag \\
	&\text{where} \quad E_v = |\{ (x,y,u) \mid (x,y,u)\ne(i,k,\ell) \,\wedge\, f_{xyu}=1 \notag \\
	&\qquad\qquad\qquad\qquad\qquad \,\wedge\, z_{xy}=z_{ik} \,\wedge\, w_{xyu}=v \}|, \notag \\
	&\text{and} \quad F_v = |\{ (x,y,u) \mid (x,y,u)\ne(i,k,\ell) \,\wedge\, f_{xyu}=0 \,\wedge\, w_{xyu}=v \}|. \notag
\end{flalign}
\normalsize
$E_v$ is the number of non-background words $\wv=v$ assigned to topic $z_{ik}$, excluding $w_{ik\ell}$.
$F_v$ is similar, but for background words (regardless of topic indicator $z$).

Ignoring the normalizer,
the distribution of $f_{ikl}$ contains a posterior term for $w_{ik\ell}$ and a prior term for $f_{ikl}$. Again, the $w_{ik\ell}$
term is a DCM; this DCM comes from integrating $\beta_{z_{ik}}$ if $f_{ik\ell}=1$, otherwise it
comes from integrating the background word distribution $\beta_{back}$.
The $f_{ik\ell}$ prior is a simple $\mathrm{Bernoulli}(\delta)$.
As with Eq. (\ref{eq:gs_z}), the counts $E_v,F_v$ can be cached with constant time updates per $f_{ik\ell}$,
thus sampling all $\fv$ is linear time in the number of words $\wv$.

\paragraph{Link topic indicators $\sv$}
Recall that we only model $s_{ij},s_{ji},e_{ij}$ for positive links
$e_{ij}=1$. For convenience, let $e_{ji}=e_{ij}$ for all $i<j$. The resulting conditional distribution of $s_{ij}$ is
\small
\begin{flalign}
	&\Pb(s_{ij} = m \mid \zv, \fv, \wv, \sv_{-ij}, e_{ij} = 1, \ev_{-ij}, \yv) \label{eq:gs_s} \\
	&\propto
		\Pb(y_{i} \mid \zv_{i}, s_{ij} = m, \sv_{i,-j})
		\Pb(e_{ij} = 1 \mid s_{ij} = m, s_{ji}, \sv_{-\{ij,ji\}}, \ev_{-ij}) \notag\\
	&\qquad\times \Pb(s_{ij} = m \mid \zv_{i}, \sv_{i,-j}) \notag \\
	&\propto
		\exp\left\{-\frac{(y_i - \hat{\theta}^{\top}_{i}\nu)^2}{2\sigma^2}\right\}
		\frac{\lambda_1 + C}{\lambda_1 + \lambda_0 + C}
		\left( \#[\{\zv_{i},\sv_{i,-j}\} = m] + \alpha \right), \notag \\
	&C = \begin{cases}
			|\{ (x,y)\in\mathrm{EdgeList} \mid (x,y)\ne(i,j) \,\wedge\, [(s_{xy},s_{yx})=(m,s_{ji}) \\
			\qquad \vee (s_{xy},s_{yx})=(s_{ji},m)] \}|
				\qquad \text{if $i<j$} \\
			|\{ (y,x)\in\mathrm{EdgeList} \mid (x,y)\ne(i,j) \,\wedge\, [(s_{xy},s_{yx})=(m,s_{ji}) \\
			\qquad \vee (s_{xy},s_{yx})=(s_{ji},m)] \}|
				\qquad \text{if $i>j$.}
		\end{cases} \notag
\end{flalign}
\normalsize
$C$ is the number of positive links $\ev \setminus e_{ij}$ whose topic indicators $(s_{xy},s_{yx})$ are identical to the topics $(s_{ij},s_{ji})$ of $e_{ij}$.
The OR clauses simply take care of situations where $s_{xy} > s_{yx}$ and/or $s_{ij} > s_{ji}$.
The distribution of $s_{ij}$ contains a prior term for $s_{ij}=m$ (the DCM from integrating $\theta_i$), a Gaussian posterior term for $y_i$,
and a link posterior term for $e_{ij}$ (the Beta Compound Bernoulli distribution from integrating out the link probability $\Phi_{m,s_{ji}}$). 

Like Eq. (\ref{eq:gs_z},\ref{eq:gs_f}), $C$ can be cached using constant time updates per $s_{ij}$, thus sampling all $\sv$
is linear in the number of friendships $|\mathrm{EdgeList}|$. Combined with the constant time sampling for Eq. (\ref{eq:gs_z},\ref{eq:gs_f}),
we see that the \SMF~algorithm requires linear time in the amount of data.

\subsubsection{Learning tuning parameters $\alpha,\eta,\delta$ and $\nu,\sigma^2$}

We automatically learn the best tuning parameters $\alpha,\eta,\delta$ using Independence Chain Metropolis-Hastings,
by assuming $\alpha,\eta$ are drawn from $\mathrm{Exponential}(1)$, while $\delta$ is drawn
from $\mathrm{Beta}(1,1)$.
For $\nu,\sigma^2$, we take a Stochastic Expectation-Maximization \cite{gilks1996markov} approach, in which
we maximize the log-likelihood with respect to $\nu,\sigma^2$ based on the current Gibbs sampler
values of $\zv,\sv$. The maximization has a closed-form solution similar to sLDA \cite{blei2008supervised}, but without the expectations:
\begin{flalign}
	&\nu \leftarrow \left( A^\top A \right)^{-1} A^\top b,
	\qquad \sigma^2 \leftarrow \frac{1}{P} \left[ b^\top b - b^\top A \nu \right] \label{eq:nusigma_mle}
\end{flalign}
where $A$ is a $P\times K$ matrix whose $i$-th row is the current Gibbs sample of $\hat{\theta}_{i}$,
and $b$ is a $P$-vector of user labels $y_i$.

Updating all parameters $\alpha,\eta,\delta,\nu,\sigma^2$ requires linear time in the amount of data, so
we update them once per Gibbs sampler sweep over all latent variables $\zv,\fv,\sv$. This ensures that every iteration
(Gibbs sweep plus parameter update) takes linear time.


\subsection{Parallelizable Prediction Algorithm}

\begin{algorithm}[t]
\caption{\SMF~Training Algorithm}
\label{alg:sm4_training}
\begin{algorithmic}[1]
\STATE {\bf Input:} Training user text data $\wv$, links $\ev$ and labels $\yv$
\STATE Randomly initialize $\zv,\fv,\sv$ and parameters $\alpha,\eta,\delta,\nu,\sigma^2$
\STATE Set $\lambda_1,\lambda_0$ according to Section \ref{sec:generative_process}, Network Model
\REPEAT
	\STATE Gibbs sample all $\zv,\fv,\sv$ using Eqs. (\ref{eq:gs_z},\ref{eq:gs_f},\ref{eq:gs_s})
	\STATE Run Metropolis-Hastings on tuning parameters $\alpha,\eta,\delta$ 
	\STATE Maximize parameters $\nu,\sigma^2$ using Eq. (\ref{eq:nusigma_mle})
\UNTIL{Iteration limit or convergence}
\STATE {\bf Output:} Sufficient statistics for $\zv,\fv,\sv$, and all parameters $\alpha,\eta,\delta,\lambda_1,\lambda_0,\nu,\sigma^2$
\end{algorithmic}
\end{algorithm}

\begin{algorithm}[t]
\caption{\SMF~Parallelizable Prediction Algorithm}
\label{alg:sm4_predict}
\begin{algorithmic}[1]
\STATE {\bf Input:} Parameters $\beta,\Phi,\alpha,\delta,\nu,\sigma^2$ from training phase
\STATE {\bf Input:} Test user $p$'s text data $\wv_p$
\STATE Randomly initialize $\zv_p,\fv_p$ for the test user
\REPEAT
	\STATE Gibbs sample $\zv_p$ using Eq. (\ref{eq:gs_z}), and $\fv_p$ using Eq. (\ref{eq:gs_f})
\UNTIL{Iteration limit or convergence}
\STATE Estimate test user's feature vector $\theta_p$ from his $\zv_p$
\STATE Use $\theta_p$ to predict $s_{pj},s_{jp}$ for all friends $j$
\STATE {\bf Output:} Test user's $\theta_p,s_{pj},s_{jp}$
\end{algorithmic}
\end{algorithm}

Our training algorithms learns topic parameters $\beta,\Phi,\nu$, so that we can use
our Prediction Algorithm (\ref{alg:sm4_predict}) to predict feature vectors $\theta_p$
and friendship topic-pair assignments $s_{pj},s_{jp}$ for {\it all} users $p$.
For each user $p$ independently and {\it in parallel}, we Gibbs sample her text latent variables $z_{p\cdot},f_{p\cdot\cdot}$
based on her observed documents $w_{p\cdot\cdot}$ and the learnt parameters $\beta,\Phi,\nu,\sigma^2$.
Then, using the definition of our \SMF~generative process, we estimate $p$'s feature vector $\theta_p$ by averaging over her $z_{p\cdot}$.
Finally, we use $\theta_p$ and the learnt topic parameters $\Phi$ to predict $p$'s most likely
friendship topic-pair assignments $s^{*}_{pj},s^{*}_{jp}$ to each of her friends $j$, using this equation:
\begin{align}
(s^{*}_{pj},s^{*}_{jp}) &= \arg \max_{(a,b) \text{ s.t. } a<b} \theta_{p,a} \Phi_{a,b} \theta_{j,b}. \label{eq:s_pred}
\end{align}
We use these assignments to discover the topics that friendships are most frequently associated with.
Like the training algorithm, the Prediction Algorithm also runs in linear time.


\section{Experimental setting}
\label{sec:experimental_setting}

Our goal is to analyze Facebook users in the context of their interests, friendships and conversations. 
Facebook users typically express interests such as ``movies" or ``cooking" by establishing a ``like" relation
with the corresponding Facebook pages, and our experiments focus on four popular user interests in Facebook:
camping, cooking, movies and sports. We selected these concepts because of their broad scope: not only are they
generic concepts, but each of their pages was associated with more than 5 million likes as of May 2011,
ensuring a sufficiently large user base for data collection.
For each interest $C$, we collected our data as follows:
\small
\begin{enumerate}
\setlength{\parskip}{0pt}
	\item
	Construct the complete data collection $S(C)$ by randomly selecting 1 million users who like interest $C$ ($y_i=+1$),
	and 1 million who do not explicitly mention liking $C$ ($y_i=-1$).
	\item
	For each user $i\in S(C)$, collect the following data\footnote{We use only non-private user data for our experiments, e.g.
	chat logs or user messages are never looked at.}:
	\begin{list}{\labelitemi}{\leftmargin=1em}
		\item
		{\bf User text documents $w_{ik\cdot}$}: The text documents for user $i$ contain
		all of her ``status updates" from March 1st to 7th, 2011 (each status update is
		one document), as well as titles of Facebook pages that she likes by March 7th 2011 (each page title is
		one document)\footnote{We remove the page title of concept $C$, because its distribution is
		highly correlated with the labels.}. We preprocessed all documents using typical
		NLP techniques, such as stopword removal, stemming, and collocation identification \cite{krenn2000collocation}.
		\item
		{\bf User-to-user friendships}: We obtained these symmetric friendships using the friend lists of user $i$ recorded on March 7th 2011.
	\end{list}
	\item
	Randomly sample $2\%$ of $S(C)$ to construct a 40,000-user training collection $\bar{S}(C)$. Across the four concepts, $\bar{S}(C)$ contained 340,128 to 385,091 unique words, 6,650,335 to 8,771,298 documents, 16,421,601 to 22,521,507 words, and 1,292 to 2,514 links\footnote{The relatively small number of links arises from unbiased random sampling of users; more links can be obtained by starting with a seed set of users and picking their friends, but this introduces bias. Also, our method uses evidence from negative links, so the small number of positive links is not necessarily a drawback.}.
\end{enumerate}
\normalsize
We first trained the \SMF~model using the training collection $\bar{S}(C)$ and $K\!=\!50$ latent features (topics),
stopping our Gibbs sampler at the 100th iteration because 1) the per-iteration increase in log-likelihood was $<\!1\%$
of the cumulative increase, and 2) more iterations had negligible impact on our validation experiments.
This process required ~24 hours for each concept, using one computational thread.
We note that one could subsample larger training collections $\bar{S}(C)$, thus increasing the accuracy of parameter learning
at the expense of increased training time. A recently introduced alternative is to apply approximate parallel inference techniques
such as distributed Gibbs sampling \cite{mimno2007organizing,asuncion2008asynchronous}, but these introduce
synchronization and convergence issues that are not fully understood yet.

After learning topic parameters from the training collection $\bar{S}(C)$, we invoke Algorithm \ref{alg:sm4_predict}
on all users $p\in S(C)$ to obtain their predicted feature vectors $\theta_p$, and the friendship
topic-pair ``explanations" $s_{pj},s_{jp}$ for each of $p$'s friends $j$. Note that Algorithm \ref{alg:sm4_predict} is
parallelizable over every user in $S(C)$, and we observe that it only requires a few minutes per user; a
sufficiently large cluster finishes all 2M users in a single day --- in fact, given enough computing power, it is possible
to scale our prediction to all of Facebook. In the following sections, we shall apply the predicted
$\theta_p,s_{pj},s_{jp}$ to various analyses of Facebook's data.

\section{Validation}

Before interpreting our results, we must validate the performance of our \SMF~model and algorithm. Because our model
spans multiple data modalities, there is arguably no single task or metric that can evaluate all aspects of \SMF. What
we shall do is test how well the \SMF~latent space and feature vectors predict held-out user interest labels $y_p$ from our data collections $S(C)$.
We believe this is the best task for several reasons: for one, we are concerned with interpreting user interests in the
context of friendships and conversations, thus we must show that the \SMF~latent space accurately captures user interests. For another,
predicting user interests is a simple and well-established task, and its results are therefore easier to interpret than model goodness-of-fit measures
such as perplexity (as used in \cite{Blei:2003:LDA:944919.944937}).

It is well-understood that textual latent space methods like Latent Dirichlet Allocation (LDA), while useful for summarization and
visualization, normally do not improve classification accuracy --- in fact, with large amounts of training data, they may actually perform worse than a naive
Bag-of-Words (BoW) representation~\cite{Blei:2003:LDA:944919.944937}. This stems from the fact that latent space methods
are dimensionality reduction techniques, and thus distort the data by necessity. In our case, the picture is more complicated:
the text aspect of our model loses information with respect to BoW, yet some non-textual information comes into play from the friendship links
and labels in the small training collections $\bar{S}(C)$. We believe the best way to use \SMF~is to concatenate \SMF~features
to the BoW features --- this avoids the information loss from reducing the dimensionality of the text, while allowing the
network and label information to come into play. We expect this to yield a modest (but statistically significant) improvement in accuracy over a plain BoW baseline.

Our task setup is as follows: recall that for each interest $C$, we obtained a 2M data collection $S(C)$ with ground truth labels for all user interests $y_p$.
The \SMF~algorithm predicts feature vectors $\theta_p$ for all users $p\in S(C)$, which can be exploited
to learn a linear Support Vector Machine (SVM) classifier for the labels $y_p$. More specifically, we use $\theta_p$ concatenated
with user $p$'s original BoW as feature inputs to LIBLINEAR~\cite{fan2008liblinear}, and then performed 10-fold cross-validation experiments on the labels $y_p$.
This was done for each of the four data collections $S(C)$, and each experiment took $<1$ hour. As a baseline, we compare to LIBLINEAR trained on
BoW features only. The BoW features for user $p$ are just the normalized word frequencies over all her documents.

Table \ref{tab:liblinear_results} summarizes our results. To
determine if the improvement from \SMF~is statistically significant, we conducted a $\chi^2$-test (one degree
of freedom, 2M trials) against the BoW Baseline as a null hypothesis. The $p$-values are far below $0.001$, suggesting that the improvement
provided by \SMF~features is statistically very significant. This confirms our hypothesis that the \SMF~features improve classification accuracy,
by virtue of encoding network and label information from the small training collections $\bar{S}(C)$. We expect that classification
accuracy will only increase with larger training collections $\bar{S}(C)$, albeit at the expense of more computation time.

\begin{table}[t]
\caption{\small User interest classification accuracy (in percent) under a 10-fold cross-validation setup,
for a Bag-of-Words baseline, and BoW plus \SMF~feature vectors. Each experiment is performed over 2 million users.
We also report $\chi^2$-statistics and $p$-values (1 degree of freedom), which show that adding \SMF~features yields a highly significant improvement in
accuracy.}
\label{tab:liblinear_results}
\small
\begin{center}
\begin{tabular}{l||r|r|r|r}
{\bf Features} &Sports &Movies &Camping &Cooking \\
\hline
\hline
BoW Baseline &78.91 &78.51 &79.85 &77.22 \\
\bf{Plus \SMF} &\bf{80.23} &\bf{80.48} &\bf{81.08} &\bf{78.57} \\
\hline
\hline
$\chi^2$-statistic & $2.1 \times 10^5$ & $4.6 \times 10^5$ & $1.9 \times 10^5$ & $2.1 \times 10^5$ \\
$p$-value & $\ll 0.001$ & $\ll 0.001$ & $\ll 0.001$ & $\ll 0.001$
\end{tabular}
\end{center}
\end{table}

\section{Understanding User Interests and Friendships in Facebook}
\label{sec:user_interest_study}

\begin{figure*}
	\vspace{-0.7cm}
	\begin{center}
	\includegraphics[width=0.9\textwidth]{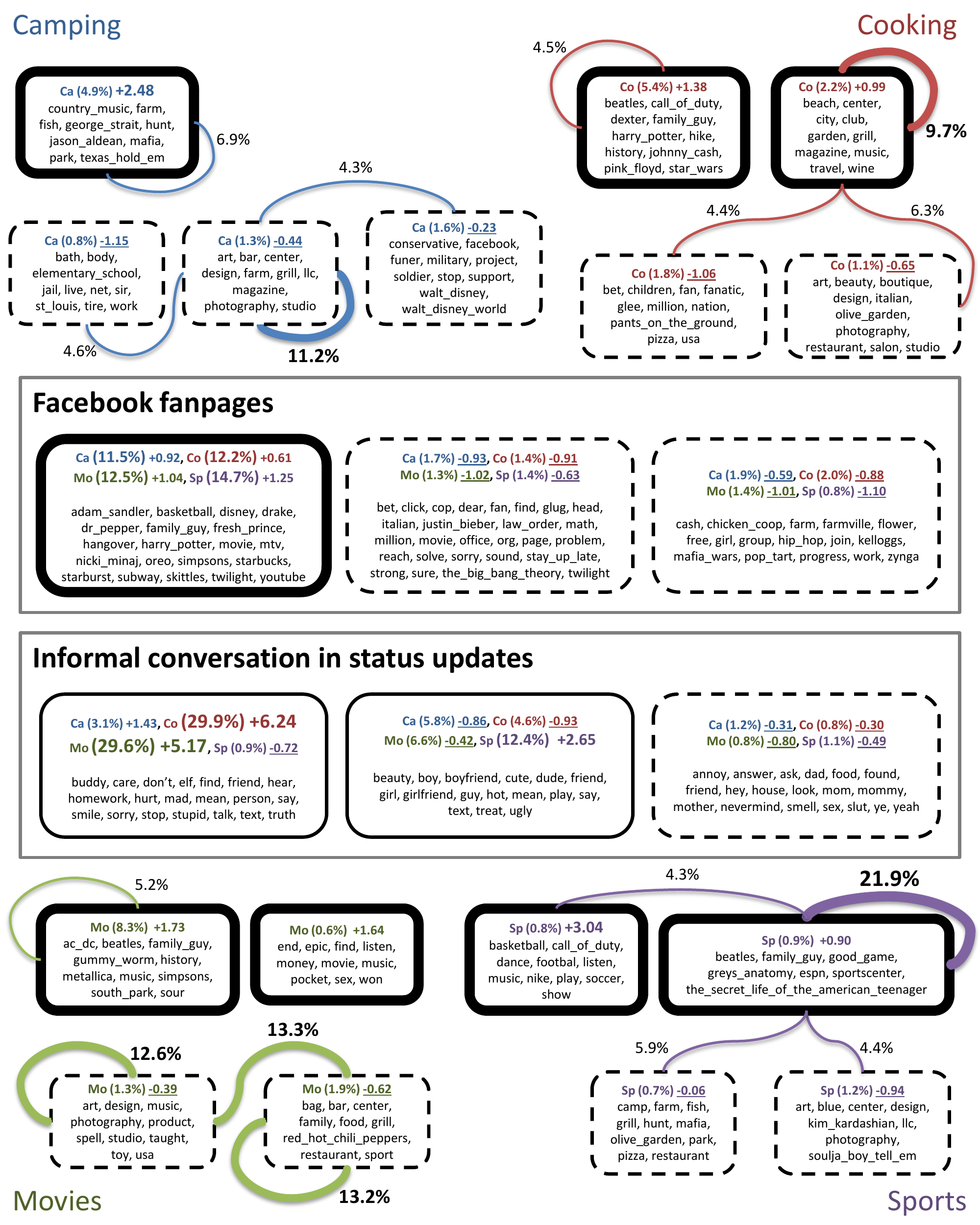}
	\end{center}
	\vspace{-0.6cm}
	\caption{\footnotesize A visual summary of the relationship between Facebook friendships, user conversations, and 4 types of user interests (best viewed in color).
	Topics specific to a particular interest are found in the corners, while common topics are found in
	the middle, divided into topics containing Facebook fanpage titles or status update lingo --- note that we manually introduced this distinction
	for the sake of visualization; the \SMF~algorithm discovers all topics purely from the data. Thick borders
	highlight topics positively correlated with user interests, while dashed borders highlight negative correlation.
	Font colors highlight information relevant to a specific interest: blue for camping (ca), red for cooking (co), green for movies (mo), and purple for sports (sp).
	The colored heading in each topic describes its popularity, and its correlation with user interests: for example,
	``Ca ($4.9\%$) $+2.48$'' means this topic accounts for $4.9\%$ of user text in the camping dataset, and
	has a moderate positive correlation with interest in camping. Finally, an edge between a pair of topics shows
	the proportion	of friendships attributed to that pair (normalized by topic popularity).}
	\label{fig:interest_viz}
\end{figure*}

In the introduction, we posed four questions about Facebook:
\small
\begin{list}{\labelitemi}{\leftmargin=1em}
\itemsep -1pt
\item
How does Facebook's social (friendship) graph interact with its interest graph and conversational content? Are they correlated?
\item
What friendship patterns occur between users with similar interests?
\item
Do users with similar interests talk about the same things?
\item
How do different interests (say, camping and movies) compare? Do groups of users with distinct interests also exhibit
different friendship and conversational patterns?
\end{list}
\normalsize
We shall answer these questions by analyzing our \SMF~output over the four user interests: camping, cooking,
movies and sports. Such analysis is not only useful for content recommendation, but can also inform policies targeted at increasing
connectivity (making more friends) and interaction (having more conversations) within the social network.
Through continuous study of user interests, conversations and friendships, we hope to learn what makes the social network unique,
and what must be done to grow it.

\subsection{Visualization procedure}

In Figure \ref{fig:interest_viz}, we combine \SMF's output over all four user interests into one holistic visualization,
and the purpose of this section is to describe how we constructed said visualization. First, recall that for each interest $C$, our \SMF~system
learns topic parameters from a training subset $\bar{S}(C)$ of user text documents,
friendship links, and labels. These parameters are then used to infer various facts about the full user dataset $S(C)$:
(1) user feature vectors $\theta_p$ that give their propensities towards various topics, and
(2) each friendship's most likely topic-pair assignments $s_{ij},s_{ji}$, which reveal the topics a given pair of friends is most likely to talk about.

With these learnt parameters, we search for the 6 most strongly-recurring topics across all four interests, as measured by cosine similarity.
These topics, shown in the middle of Figure \ref{fig:interest_viz}, represent commonly-used words on Facebook, and provide a
common theme that unites the four user interests. Next, for each interest, we search for the top 4 topic-pairs (including pairs of the same topic)
with the highest friendship counts (which come from the topic-pair assignments $s_{ij},s_{ji}$). Note that we first normalize each topic-pair friendship
count by the popularity\footnote{The sum of a topic's weight over all user feature vectors $\theta_p$.} of both topics,
in order to avoid selecting popular but low-friendship topics. We show these 4 topic-pairs in the corners of Figure \ref{fig:interest_viz}, along
with their normalized friendship counts. These topic-pairs represent conversations between friends; more importantly, if the topics
are also positively correlated with the user interest --- say, camping --- then they reveal what friends who like camping actually talk about.
This context-specificity is especially valuable for separating generic chatter from genuine conversation about an interest.

Figure \ref{fig:interest_viz} was constructed by these rules, but with one exception: we include a Movies topic (heading
Mo ($0.6\%$) $+1.64$) that lacks strong friendships, yet is positively correlated with interest in movies. This anomaly demonstrates
that interest-specific conversations do not always occur between friends --- in other words, the presence of an interest-specific conversation
does not imply the existence of friendship, which is something that text-only systems may fail to detect. In turn,
this highlights the need for holistic models like \SMF~that consider interests, conversations and friendships jointly.

\subsection{Observations and Analysis}

\paragraph{Common Topics}

Throughout these sections, we shall continually refer to Figure \ref{fig:interest_viz}. The most striking observation about the four interests (camping,
cooking, moving, sports) is their {\it shared topical content}, shown in the middle of the Figure. These topics represent a common lingo
that permeates throughout Facebook, and that can be divided into two classes: ``Facebook fanpages", consisting of named entities
that have pages on Facebook for users to like, and "Informal conversation in status updates", which encompasses the most common, casual words from user status updates.

We observe that the fanpage topic starting with ``adam\_sandler'' is dominant, with popularity  $>\!10\%$ across all four user interest datasets.
Additionally, this topic has a mild positive correlation with all interests, meaning that users who have any of the four interests are more likely to use this topic. In
contrast, the fanpage topic starting with ``cash" only has average popularity (between $1-2\%$) and mild negative correlation with all interests. Observe that this topic is
dominated by social gaming words (``farmville", ``mafia\_wars"), whereas the other, popular topic is
rich in popular culture entities such as ``Disney", ``Dr Pepper", ``Simpsons" and ``Starbucks". This data provides evidence that users who exhibit any of the four
interests tend to like pop culture pages over social gaming pages. Notably, none of these four interests are related to internet culture or gaming,
which might explain this observation.

The informal conversation topics are more nuanced. Notice how the topic starting with ``buddy" is both popular and strongly correlated with
respect to cooking and movies, implying that the conversations of cooking/movie lovers differ from camping/sports lovers.
Also, notice that the topic starting with ``beauty" is dominated by romantic words such as ``boyfriend" and ``girlfriend", and is popular/correlated only with sports ---
perhaps this lends some truth to the stereotype that school athletes lead especially active romantic lives. Finally, the topic starting with ``annoy"
and containing words such as ``dad", ``mom" and ``house" carries a slight negative sentiment for all interests (in addition to being unpopular).
This seems reasonable from the average teenager's perspective, in which parents normally have little connection with personal interests.

\paragraph{High-Friendship Topics}

We turn to the high-friendship topics in the corners of Figure \ref{fig:interest_viz}. Some of these
contain a high degree of self-friendships, implying that friends usually converse about the same topic, rather than different ones.
To put it succinctly, in Facebook, the interest graph is correlated with the social (friendship) graph.
In fact, the average proportion of same-topic friendships ranges from $0.2\%$ to $0.6\%$ depending on interest, whereas the average proportion of inter-topic
friendships is an order of magnitude lower at $0.02\%$ to $0.04\%$. Intuitively, this makes sense: any coherent dialogue between friends 
is necessarily about a single topic; multiple-topic conversations are hard to follow and thus rare.

One interpretation of inter-topic friendships is that they signify two friends who rarely interact, hence their conversations on the whole are topically distinct.
In other words, inter-topic friendships may represent socially weaker ties, compared to same-topic friendships. As an example, consider the cooking topics starting with
``art" and ``conservative" respectively. The former topic is about the visual arts (``design", ``photography", ``studio"), whereas
the latter topic is about political conservatives in America (``military", ``soldier", ``support"). It seems implausible
that any conversation would be about both topics, and yet there are friendships between people who talk about either topic --- though not necessarily
with each other.

A second observation is that most interests have more than one positively correlated topic (with the exception of camping). A good
example is cooking: notice the topics starting with ``beach" and ``beatles" respectively. The former topic has connotations of fine living, with
words like ``city", ``club", ``travel" and ``wine", whereas the latter is associated with entertainment culture, containing phrases like ``beatles",
``family\_guy", ``pink\_floyd" and ``star\_wars". Both topics have statistically much in common: moderate popularity, positive interest correlation with cooking,
and a significant proportion of self-topic friendships. Yet they are semantically different, and more importantly, do not have a significant proportion of friendships
between them. Hence, these two topics represent separate communities of cooking lovers: one associated with the high life, the other with pop culture.
The fact that cooking lovers are not homogenous has significant implications for policy and advertising; a one-size-fits-all strategy is unlikely to succeed.

Similar observations can be made about sports and movies: for sports, both a television topic
(``family\_guy", ``greys\_anatomy", "espn") and an actual sports topic (``basketball", ``football", ``soccer") are
positively correlated with interest in sports, yet users in the former topic are likely {\it watching} sports rather than {\it playing} them.
As for movies, one topic is connected with restaurants and bars (``bar", ``food", ``grill", ``restaurant"), while the other is connected
with television (``family\_guy", ``simpsons", ``south\_park").

Our final observation concerns the ``friendliness" of users in positive topics --- notice that the users of some positively correlated topics (``country\_music" from camping,
``ac\_dc" from movies, ``beatles" from cooking") have plenty of within-topic friendships, yet possess almost no friendships with other topics. In contrast,
users in topics like ``beach" from cooking or ``beatles" from sports are highly gregarious, readily making friends with users in other topics. The
topic words themselves may explain why: notice that the ``beach" cooking topic has words like ``club", ``grill" and ``travel" that suggest highly social activities,
while the ``beatles" sports topic contains television-related words such as ``family\_guy" and ``espn", and television viewing is often a social activity as well.

In closing, our analysis demonstrates how a multi-modal visualization of Facebook's data can lead to insights about network connectivity
and interaction. In particular, we have seen how fanpages and casual speech serve as a common anchor to all conversations on Facebook,
how same-topic friendships are far more common (and meaningful) than inter-topic friendships, and how users with common interests can be hetorogenous
in terms of conversation topics. We hope these observations can inform policy directed at growing the social network, and increasing the engagement of its users.

\section{Related Work and Conclusion}

The literature contains other topic models that combine several data modalities; ours is distinguished by the assumptions it makes.
In particular, existing topic models of text and network data either treat the
network as an outcome of the text topics (RTM \cite{chang2009relational}), or define new topics for each link
in the network (ART \cite{mccallum2007topic}). The Pairwise Link-LDA model of Nallapati {\it et al.}~\cite{nallapati2008joint} is the most similar to ours,
except (1) it does not model labels, (2) it models asymmetric links only, and crucially, (3)
its inference algorithm is infeasible for even $P=40,000$ users (the size of our training $\bar{S}(C)$'s) because
it models all $\mathrm{O}(P^2)$ positive {\it and} zero links. Our model escapes this complexity trap by only
considering the positive links.

We also note that past work on Facebook's data \cite{wang2011learning} used
the network implicitly, by summing features over neighboring users.
Instead, we have taken a probabilistic perspective, borrowing from the MMSB model \cite{airoldi2008mixed}
to cast links into the same latent topic space as the text.
Thus, links are neither a precursor to nor an outcome of the text, but {\it equals},
resulting in an intuitive scheme where both text and links derive from specific topics.
The manner in which we model the labels is borrowed from sLDA \cite{blei2008supervised}, except that
our links also influence the observed labels $\yv$.

In conclusion, we have tackled salient questions about user interests and friendships on Facebook,
by way of a system that combines text, network and label data
to produce insightful visualizations of the social structure generated by millions of Facebook users.
Our system's key component is a latent space model (\SMF) that learns the aggregate relationships between
user text, friendships, and interests, and this allows us to study millions of users at a macroscopic level.
The \SMF~model is closely related to the supervised text model
of sLDA \cite{blei2008supervised} and the network model of MMSB \cite{airoldi2008mixed},
and combines features of both models to address our challenges.
We ensure scalability by splitting our learning algorithm into two phases:
a training phase on a smaller user subset to learn model parameters, and a parallel prediction
phase that uses these parameters to predict the most likely topic vectors $\theta_p$ for each user,
as well as the most likely friendship topic-pair assignments $s_{ij},s_{ji}$ for all friendships $e_{ij}=1$.
Because the inference phase is trivially parallelizable, our system potentially scales to all users in Facebook.


\bibliographystyle{abbrv}
\bibliography{references}

\end{document}